# QUANTUM-MECHANICAL WAVES
# IN CLOSED VACUUM STATES

Paul S. Wesson

1. Dept. Physics and Astronomy, University of Waterloo, Waterloo, Ontario N2L 3G1, Canada.

2. Herzberg Institute, National Research Council, Victoria, B.C. V9E 2E7, Canada.

Abstract: Campbell's theorem enables the embedding of 4D anti-deSitter space in 5D canonical space, so a particle becomes a wave in the extra dimension, running through spacetime. This model of wave-particle duality provides a new approach to particle mass.

1. <u>Introduction</u>

In order to better understand the properties of particles, Dirac in 1935 considered embedding the simplest curved 4D space (deSitter) in the basic flat 5D space (Minkowski) [1]. The results were interesting, but did not lead to a significantly deeper understanding of particle mass. However, $deS_4$ with both positive and negative values of the cosmological constant $\Lambda$ can also be embedded in the five-dimensional canonical space $C_5$, which was introduced in 1994 [2]. As will be shown below, $C_5$ with $\Lambda < 0$ leads automatically to wave-particle duality, with a straightforward geometrical explanation of mass.

Campbell's theorem lies behind much recent work on higher-dimensional gravity. While space-time-matter theory and membrane theory were developed largely from the physical side, an algebraic side was added by the rediscovery of this theorem in 1995 and its subsequent reproof using modern techniques [3]. It is particularly relevant to the canonical metric, wherein the 5D interval consists of a quadratic factor in the extra coordinate (divided by a physical length scale) multiplied onto 4D spacetime, plus an extra flat dimension (see below). Physically $C_5$ and $M_5$ are akin to how a 3D problem may be described in either polar or Cartesian coordinates. When the 4D subspace does not depend on the extra coordinate (except via the quadratic prefactor), there results the pure form of the canonical metric $C_5^*$. This has special importance, because following Campbell it is a theorem that *any vacuum solution of 4D general relativity may be locally embedded in* $C_5^*$ [4]. This applies to the Schwarzschild solution of Einstein's equations, and to the deSitter solution. But whereas in the former case the 5D metric is not flat, in the latter case it *is* flat. Indeed, the embedding of $deS_4$ space in $M_5$ is well understood [5]. This is important for what follows, because it means that any singularities which appear in other 5D embeddings of $deS_4$ must be due to the coordinates and not be geometrical in nature. It will become clear below that the use of $C_5^*$ for the embedding of $deS_4$ leads directly to wave-particle duality with quantization, a correspondence which is obscured by the use of $M_5$ as in previous work [1, 4, 5]. That is, the embedding outlined below in Section 2 leads to new physics. It is not, in particular, equivalent to embeddings of 4D in 5D which use separable coordinates, as pioneered by Ponce de Leon [6] and which include the cosmological form of the deSitter solution. Neither is it the same as the old approach of Wesson [7], who attempted to use the fifth coordinate to give a Machian description of particle mass. (It is an algebraic irony that the form of the prefactor in the canonical metric allows of equal physical support for either of two alternatives, namely that mass is related to the extra coordinate and measured by the Schwarzschild radius, or that it is related to the length scale associated with that coordinate and is measured by the Compton wavelength, where the second hypothesis is used below.) It will not be necessary to employ Campbell's theorem directly in what follows. However, it should be noted that the 5D signature affects the sign of the 4D cosmological constant [2-7]. Specifically, a spacelike extra coordinate implies $\Lambda > 0$ while a timelike extra coordinate implies $\Lambda < 0$. Both choices are algebraically acceptable; and physically there is no problem with the closed timelike curves of 4D relativity [8], because the fifth dimension does not have the nature of time. Rather the extra variable should be viewed in a quantum-mechanical sense as the coordinate conjugate to the mass.

In Section 2, the focus is on 5D pure-canonical spaces with signature $(+---+)$. Accordingly, the 4D hypersurface $s$ of spacetime is closed, and the effective 4D vacuum is measured by a cosmological 'constant' $\Lambda$ which is negative [2, 4, 5]. The application is therefore to particles and their vacuum fields, which imply a large magnitude for $\Lambda$, rather than to cosmology with a small value of $\Lambda$ [9, 10]. The difference is, of course, well known in the form of the cosmological-'constant' problem. It would be inappropriate to rediscuss the details of that problem here; except to point out that in most 5D approaches, the cosmological 'constant' is actually a variable parameter whose value depends on the hypersurface on which it is observed [4, 6]. This generic feature of 5D field theory holds also in the model to be examined.

The notation is standard. Lower-case Greek letters run 0, 123 for time and space. Upper-case Latin (i.e. English) letters run 0, 123, 4 with the extra coordinate labelled $x^4 = l$ to avoid confusion with other applications. The speed of light ($c$), quantum of action ($h$) and gravitational constant ($G$) are set to unity, except where they are made explicit to aid physical understanding.

2. <u>5D Canonical Space and 4D deSitter Space</u>

Campbell's theorem, as noted above, ensures that Einstein's vacuum equations

$$R_{\alpha\beta} = \Lambda g_{\alpha\beta} \quad (\alpha, \beta = 0,123) \tag{1}$$

are locally embedded in the Ricci equations

$$R_{AB} = 0 \quad (A, B = 0, 123, 4) \quad . \tag{2}$$

These 5D equations allow of the calculation of $\Lambda = \Lambda(x^4)$ by a match to the 4D equations (1) on some hypersurface $x^4 = l$ which defines spacetime. This can be done given a form for the metric.

The 5D canonical metric which embeds 4D anti-deSitter space is

$$dS^2 = \left(\frac{l - l_0}{L}\right)^2 ds^2 + dl^2 \tag{3}$$

$$ds^2 \equiv \left(1 - \frac{\Lambda r^2}{3}\right) dt^2 - \left(1 - \frac{\Lambda r^2}{3}\right)^{-1} dr^2 - r^2 \left(d\theta^2 + \sin^2\theta d\phi^2\right) \quad , \tag{4}$$

where the spatial coordinates are spherical polars and $\Lambda < 0$. The 4D metric (4) is easily mapped to a pseudosphere with constant positive curvature $1/L^2$ [5] and it is this constant radial length which appears in the 5D metric (3). The other constant length $l_0$ which appears there sets the origin for the extra coordinate $x^4 = l$.

Extensive work has been done on canonical metrics with the form (3), which are so called because of the simplification they bring to the field equations and the equations of motion [4]. Probably the quickest way to confirm (3) is to consider the component $R_{44} = 0$ of (2) for a general metric of the form

$$dS^2 = f(l) ds^2 + g(l) dl^2 \quad . \tag{5}$$

Some algebra shows that the embedding function of $f(l)$ and the scalar potential $g(l)$ are strongly constrained in terms of each other. And since $g(l)$ in (5) may be absorbed by a coordinate transformation, the result is (3), which is unique up to such transformations. The field equations (2) are second order, and the lengths $L$, $l_0$ in (3) correspond to the two constants of integration involved in the solution.

Topology, however, is not fixed by the field equations, and requires a choice based on dynamics. It has been known for years that the motion of a massive particle on a 4D timelike path $(ds^2 > 0)$ may correspond to a 5D *null path* $(dS^2 = 0)$. That is, conventional causality in spacetime may correspond to simultaneity in higher dimensions. In (3), $dS^2 = 0$ leads to a path $l = l(s)$ which fixes the extra coordinate as a function of 4D proper time, this being the parameter usually employed in describing dynamics. However, the topology of that path requires a choice of physical model. It should be noted that the same choice is necessary if the more detailed dynamical information is employed that follows from the variational problem of (3) with $\delta\left[\int dS\right] = 0$. This condition leads to 4D geodesic motion *identical* to that in general relativity, because the spacetime section of (4) is decoupled from the extra dimension by virtue of $\partial g_{\alpha\beta}(x^\gamma, l)/\partial l = 0$. There is also an extra equation of motion for the $x^4 = l$ axis, which is satisfied by the null path of (3). Irrespective of whether the metric or the geodesic is used to obtain $l = l(s)$, its interpretation requires a choice of physical model.

Two choices are obvious. A linear model follows from (3) with $dS^2 = 0$ and $l_0 = 0$, such that $l = l_* \exp(\pm is/L)$. This is a wave with amplitude $l_*$ and wavelength $L$, propagating about $l = 0$ and along the hypersurface $s$ of spacetime. This model is acceptable but not very interesting, and neglects the fact noted above, namely that anti-deSitter space (4) can be visualized as a pseudosphere of radius $L$ with a closed topology. A circular model is therefore more relevant. In this, the wave can be chosen to propagate about the surface defined by $l_0 = L$, and *around* spacetime, as illustrated in Figure 1.

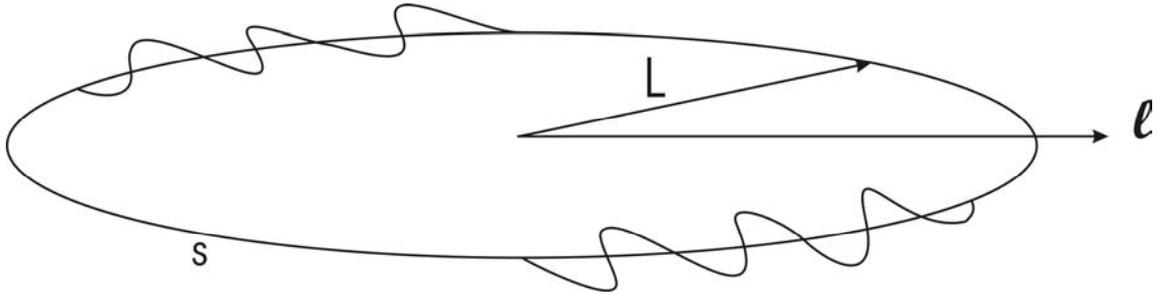

**Figure 1.** A schematic illustration of the model. Four-dimensional anti-deSitter space may be viewed as a pseudosphere of radius $L$, whose closed circumference $s$ defines spacetime. Orthogonal to this is the extra coordinate $x^4 = l$ of five-dimensional (pure) canonical space. If the 5D interval is null, the path $l = l(s)$ is a wave running around spacetime.

The circumference of spacetime $s$ is $2\pi L$, so the fundamental mode of the wave has wavelength $2\pi L$. It is natural to identify this as the Compton wavelength $1/m$ of the test particle with mass $m$ associated with the wave. Overtones or excited modes will have wavelengths shorter than the fundamental one by an integer factor $n$. Including the relevant physical constants, the wave is given by

$$l = L + l_* e^{\pm inmcs/\hbar} \quad . \tag{6}$$

This is, of course, just the description of a particle according to elementary wave mechanics. It has been derived here by applying the principle of the null path to the 5D canonical metric (3) which embeds the 4D deSitter metric (4). It is more common to see (6) quoted as the basic solution of the Klein-Gordon equation,

$$\Box^2 \psi + m^2 \psi = 0 \quad . \tag{7}$$

Here the d'Alembertian or 4D Laplace operator acts on a scalar wave function whose source is effectively the mass $m$ of the particle. Above, it was mentioned that the 5D geodesic equation for the canonical metric (3) splits into a 4D part which is identical to the geodesic equation of general relativity plus another equation of motion for the extra dimension. It can be shown by some rather tedious algebra that this latter relation is mathematically equivalent to the Klein-Gordon equation. This means that the oscillation of the extra coordinate in 5D relativity is equivalent in some sense to the wave behaviour of old quantum mechanics. Some connection of this type might have been expected from the 1936 study of Dirac [1]. However, the present model goes considerably further, as will now be shown.

The cosmological 'constant' $\Lambda$ is central to the present model, and the 5D approach helps elucidate some of the well-known problems with this parameter in lower dimensions. In 3D, there is a force (per unit mass) or acceleration $\Lambda r/3$ which acts between an arbitrarily chosen origin and a point at radius $r$, even though there may be *no* material object located at either point. This implies that $\Lambda$ is not gravitational in nature, an inference confirmed by the vacuum form of Einstein's equations (1) where the gravitational constant $G$ does not appear. In 4D, $\Lambda$ may appear explicitly, or be included implicitly as part of the energy-momentum tensor, in which case Einstein's equations read $G_{\alpha\beta} = \left(8\pi G/c^2\right)T_{\alpha\beta}$. The vacuum fluid then has a density and pressure given by $\rho_v c^2 + p_v = 0$, where $\rho_v \equiv \Lambda c^2/8\pi G$. However, this is a ruse: the $G$ in the Einstein coupling constant exactly cancels the $G$ in the vacuum density, so as before the $\Lambda$ term is seen to be non-gravitational in nature. The 4D deSitter solution describes essentially how *vacuum* curves spacetime. This is why it has been applied to quantum problems, such as tunnelling, though the values required are up to $10^{120}$ times larger than those inferred from cosmology. For $\Lambda < 0$, a test particle cannot move arbitrarily far from the origin in 3D, and the topology is a closed sphere in 4D (see above). Viewed from 5D, a particle in spacetime necessarily describes a closed orbit, given by (6) when the metric is the 5D canonical one (3). In the 5D approach, $\Lambda$ can be evaluated in two complementary ways: the intrinsic value is the one defined in the 4D hypersurface of spacetime, and uses Ein-stein's equations in the limit of no ordinary matter as in (1); while the extrinsic value is the one defined by the extra coordinate, which is orthogonal to spacetime, and uses the Ricci equation in the form (2). These values match, of course, on the spacetime hypersurface. The intrinsic value has been long known, and is $\Lambda = \pm 3/L^2$, depending on the signature of the embedding space [2, 4, 5]. The extrinsic value can be calculated by using the 5D field equations (2) to obtain the 4D Ricci or curvature scalar $^4R$. Since the canonical 5D metric (3) embeds 4D vacuum solutions like (4), the usual relation $^4R = 4\Lambda$ holds, though the result will now be of the form $\Lambda = \Lambda(l)$. For anti-deSitter, it is

$$\Lambda = \frac{^4R}{4} = -\frac{3}{(l-l_0)^2} \quad . \tag{8}$$

This agrees with the aforementioned result $\Lambda = -3/L^2$ on the hypersurface $(l-l_0) = L$ where the 5D metric (3) goes to the 4D one (4). However, it is now apparent that there is a divergence at $l = l_0$ in $\Lambda$ and $^4R$. This is reminiscent of membrane theory. But here, it should be recalled from above that since $AdeS_4$ can be smoothly embedded in $M_5$, this divergence must be merely a coordinate singularity in the 4D space, similar in nature to the horizon in the Schwarzschild metric. Nevertheless, (8) shows that in the canonical (not Minkowski) coordinates being used, the 5D metric (3) implies 4D structure as in (6) and (8). In fact, with the choice $l_0 = L$, the physics around spacetime is summed up by the wave (6) just found. And since the physics is not gravitational in nature, it becomes clear that what is involved is simply a wave in the vacuum, whose path defines spacetime and whose mechanics are characterized by $\Lambda = -3/L^2$.

Wave-particle duality is inherent to this model. The wave propagates around the circumference $2\pi L$ of spacetime, which must equal the Compton wavelength $h/mc$ of the associated particle of mass $m$. Therefore $L = \hbar/mc$, and since $|\Lambda| = 3/L^2$ there comes

$$|\Lambda| = 3(mc/\hbar)^2 \quad . \tag{9}$$

This predicts a large magnitude for the cosmological 'constant', of the sizes found in particle physics, as opposed to the small (and positive) values found in cosmology [4, 9, 10]. The complete resolution of the cosmological-'constant' problem is beyond the scope of the present model. But as an illustration, for the electron the value of $|\Lambda|$ by (9) is approximately $3 \times 10^{21} \text{cm}^{-2}$.

Quantization is also inherent to the model. This may be inferred directly from the existence of the wave (6). The amplitude $l_*$ in that relation implies an essential fuzziness to the path $l(s)$, measured by the *r.m.s.* quantity $l_*/\sqrt{2}$. But if this is small compared to the radius of the orbit $l \simeq l_0 = L$, then $dl \simeq (l - l_0)$ in the metric (3). That relation, with a trivial Wick rotation and the *null-path* condition $dS^2 = 0$, yields $ds^2/L^2 = 1$. This with $L = \hbar/mc$ (above) gives

$$mcds = \pm \hbar \quad . \tag{10}$$

This result follows mainly from the fact that the 5D metric has the canonical form (3), and would not in general follow from other 5D metrics (e.g. Minkowski; the presence of $\hbar$ rather than $h$ here reflects the choice of a circular rather than linear orbit). In other words, the (special) canonical metric $C_5^*$ leads *automatically* to the standard condition (10) for quantization.

3. Conclusion

Waves are a natural consequence of the embedding of 4D anti-deSitter space in 5D canonical space. The model examined here is the simplest possible, but some implications of it will be outlined below, after a brief review of procedure.

In Section 2, it was noted that the 4D field equations (1) can be embedded in the 5D ones (2). The latter admit 5D canonical space (3), which contains 4D deSitter space (4). The form of (3) can be proved quickly as an algebraic exercise in embedding spaces like (5). When the null-path condition is applied to 5D canonical space, the result is (6), which describes a wave oscillating in the extra coordinate about the 4D hypersurface of spacetime. Incidentally, the general equation of motion in the extra dimension of canonical space may be shown to be equivalent to the Klein-Gordon equation (7) of relativistic wave mechanics. The 4D cosmological constant $\Lambda$ has associated with it a length scale which measures the radius of anti-deSitter space considered as a pseudosphere, and the surface of this sphere is taken to be the locus of the 5D wave. The magnitude of $\Lambda$ as measured by the embedding coordinate is give by (8). This leads to a relation (9) between $\Lambda$ and the mass $m$ of the particle associated with the wave, because the fundamental wavelength of the wave is taken to be the Compton wavelength (or inverse mass) of the particle. The wave blurs the surface of spacetime, but its underlying canonical space causes automatic quantization (10) of the standard type.

Implications of the above model are both obvious and subtle. The vacuum fields of particles as measured by their effective local cosmological 'constants' should vary as the squares of their masses, a result which is robust against changes in the topology and is testable. The choice of a circular topology versus a linear one, as made above, leads to a classification scheme for particles which requires a little imagination. The 5D picture, taken in cross-section, is one of concentric orbits about a common centre. Each orbit is a spacetime. Particles of larger mass are nearer the centre in 5D, while the electron as the lightest known stable particle occupies the outermost orbit. Values of the cosmological 'constant', as measures of vacuum energy by spacetime curvature, are large in the inner regions and small in the outer regions. This picture is simple and clean as long as only vacuum is considered.

Mass and matter will, to a certain degree, muddy the picture. It is already known how to handle matter in 5D relativity [4]. A logical first step might appear to be to change the embedded 4D space from deSitter to Schwarzschild. However, while the appropriate canonical form is known, it is not 5D flat [2, 3], something which may cause algebraic difficulties. By contrast, the 4D Friedmann-Robertson-Walker spaces *are* flat in 5D, so if they were cast into canonical form they might repay investigation, especially with regard to the prevalence of a small, positive cosmological 'constant' on large scales. Of course, on small enough scales the universe *is* a vacuum, so the model outlined here has some generic value.

Acknowledgement

Thanks go to various members of the S.T.M. group, http://astro.uwaterloo.ca/~wesson.